\documentclass[12pt,preprint]{aastex}

\usepackage{underscore}
\usepackage{color}
\usepackage[dvipsnames]{xcolor}

\slugcomment{Accepted for publication in The Astronomical Journal}

\begin{document}

\title{Anisotropic Ejection from Active Asteroid P/2010 A2: \\
An Implication of Impact Shattering on an Asteroid \footnote{Based in part on data collected at Subaru Telescope, which is operated by the National Astronomical Observatory of Japan.}}

\author{Yoonyoung Kim, Masateru Ishiguro}
\affil{Department of Physics and Astronomy, Seoul National University,
Gwanak, Seoul 151--742, South Korea}
\email{yoonyoung@astro.snu.ac.kr, ishiguro@astro.snu.ac.kr}

\author{Tatsuhiro Michikami}
\affil{Faculty of Engineering, Kindai University, Hiroshima Campus, 1 Takaya Umenobe, Higashi-Hiroshima, Hiroshima 739--2116, Japan}

\author{Akiko M. Nakamura}
\affil{Department of Planetology, Kobe University, Nada, Kobe 657--8501, Japan}


\begin{abstract}
We revisited a mass ejection phenomenon that occurred in asteroid P/2010 A2 in terms of the dynamical properties of the dust particles and large fragments. We constructed a model assuming anisotropic ejection within a solid cone-shaped jet and succeeded in reproducing the time-variant features in archival observational images over $\sim$3 years from 2010 January to 2012 October. When we assumed that the dust particles and fragments were ejected in the same direction from a point where no object had been detected in any observations, the anisotropic model can explain all of the observations including (i) the unique dust cloud morphology, (ii) the trail surface brightness and (iii) the motions of the fragments. Our results suggest that the original body was shattered by an impact with the specific energy of $Q^{\ast} \la$ 350~J kg$^{-1}$, and remnants of slow antipodal ejecta (i.e., anisotropic ejection in our model) were observed as P/2010 A2.
The observed quantities are consistent with those obtained through laboratory impact experiments, supporting the idea that the P/2010 A2 event is the first evidence of the impact shattering occurred in the present main asteroid belt.

\end{abstract}
\keywords{minor planets, asteroids: general --- minor planets, asteroids: individual (P/2010 A2)}

\clearpage
\section{Introduction}
\label{sec:introduction}

\begin{figure}
\epsscale{0.7}
\plotone{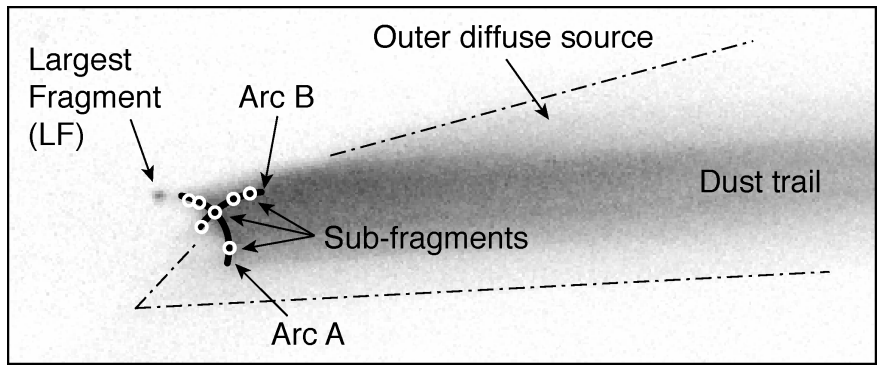}
\caption{Schematic diagram illustrating the major features of P/2010 A2 onto the background image of Gemini-2010. The image has the standard orientation in the sky: north is up and east is to the left. The FOV is $1.2\arcmin \times0.5\arcmin$. The open circles denote the positions of sub-fragments, while the thick solid lines show the locations of arc A and arc B. All of these fine structures are written in \citet{A13}. The outer diffuse source is enclosed by the dot-and-dash lines.
}
\label{fig:schematic}
\end{figure}

P/2010 A2 (LINEAR) is the fifth recognized ``active asteroid''. It was discovered on 2010 January 6 by the Lincoln Near Earth Asteroid Research (LINEAR) survey \citep{discovery}. It orbits in the main asteroid belt but exhibited a dust ejection like comets \citep{HJ06,J15}.
After the discovery, intensive follow-up observations were performed using a variety of ground-based and space telescopes to reveal the mass ejection mechanism.  It displayed a distinctive morphology of the dust cloud (see Figure \ref{fig:schematic}) with a prominent point-like source that is approximately 120~m in diameter (hereafter, the largest fragment, LF) at the leading edge of the dust trail. Two arc-like structures were noticed at the eastern edge of the dust trail (arc A and B). In addition, several decimeter-sized or larger sub-fragments were found along the arcs \citep{J10,A13}. It was also noticed that a fainter structure (hereafter outer diffuse source) extended more widely than the dust trail \citep{H12,K13}.

The mass ejection mechanism of the asteroid still remains inconclusive, although there are several efforts to understand the cause through dynamical modeling of the dust particles and fragments. In an early study, \citet{M10} considered that the dust particles were ejected continuously by sublimation of ice. Later, it became clear that the mass was ejected impulsively rather than continuously on a day in 2009 February or March by either an impact or rotational breakup \citep{J10}. \citet{K13} argued that the arcs are associated with an impact hollow cone, suggesting that a decameter-sized crater was formed on the LF. On the other hand, \citet{A13} studied the motions of sub-fragments using a series of high-resolution images taken by the \textit{Hubble Space Telescope} (HST) from 2010 January to May and suggested that the mass was ejected in the equatorial plane of the LF by centrifugal force (called rotational breakup). \citet{J13} further obtained observational image in 2012 October from the 10 m Keck I telescope and conducted a model simulation of dust particles to understand the observed image. They claimed that their observation was consistent with an impact close to the shattering threshold, although they could not rule out the possibility of a rotational breakup. Therefore, the interpretations to the comet-like activity are incoherent, that is, either impact shattering, cratering or rotational breakup.

\begin{figure}
\epsscale{0.9}
\plotone{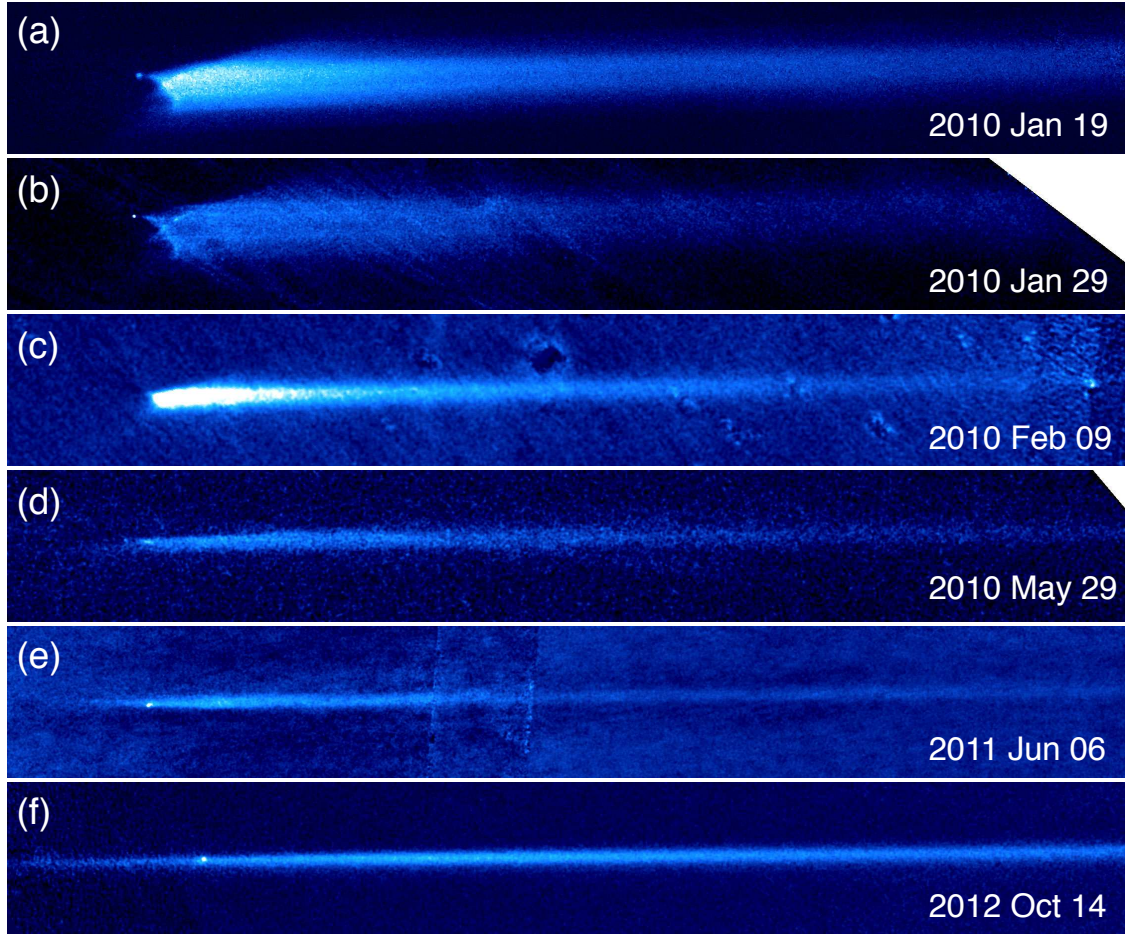} 
\caption{The observed composite images at six different epochs: (a) Gemini-2010 (138$\arcsec$ wide), (b) HST-2010 (96$\arcsec$ wide), (c) CHFT-2010 (257$\arcsec$ wide), (d) HST-2010 (73$\arcsec$ wide), (e) Subaru-2011 (521$\arcsec$ wide) and (f) Keck-2012 (310$\arcsec$ wide). These images were rotated to align the trail orientation horizontally. Note that the background objects such as galaxies and stars in each exposure were erased when we combined these images \citep[see, e.g.,][]{Ishiguro2008}.
}
\label{fig:obs}
\end{figure}

Here, we note that none of the previous dynamical modeling are successful in reproducing the multi-epoch observed features (i.e., the time-variant dust cloud morphology, the trail surface brightness and the motions of fragments). For example, \citet{J13} did not deal with observation images taken before 2012, while \citet{A13} seems to not replicate the trail surface brightness in 2012 (see Figure 8 in the reference). With the exception of \citet{A13}, no studies considered the motions of the fragments. To complement the incomplete modeling and elicit further information about this mysterious phenomenon, we revisited archival observations taken from 2010 January to 2012 October together with our unpublished observation conducted using the 8.2 m Subaru telescope in 2011 June (see Figure \ref{fig:obs}, Table \ref{tab:observation}, and the abbreviations therein) and constructed a new model that could replicate these time-variant features in the archival observation data over $\sim$3 years since its discovery.

The rest of this paper is organized as follows. Section \ref{sec:model} describes the model details, Section \ref{sec:result} presents the results of the dust modeling and further analysis of the fragment motion. We discuss the results based on the knowledge obtained through laboratory impact experiments in Section \ref{sec:discussion} and summarize the findings and their physical implications in Section \ref{sec:summary}.

\clearpage
\section{Model Description}
\label{sec:model}

To understand the time-variant morphology and surface brightness distribution over the entire period of the available observations (i.e., $\sim$3 years from 2010 January to 2012 October) in a comprehensive manner, we conducted a dynamical simulation of dust particles taking into account solar gravity and radiation pressure \citep{Ishiguro2007,I11,Ishiguro2013}. The trajectory of a dust particle is determined by the particle radius and the ejection velocity \citep{Finson1968}. The size of the particle can be parameterized by $\beta$, the ratio of radiation pressure acceleration to solar gravity. For a spherical particle, $\beta$ is given by

\begin{eqnarray}
\beta = \frac{K Q_\mathrm{pr}}{\rho_\mathrm{d} a_\mathrm{d}},
\end{eqnarray}

\noindent where $a_\mathrm{d}$ and $\rho_\mathrm{d}$ are the particle radius and the mass density in the MKS units, respectively. For large fragments, the $\beta$ values can be approximated to zero. Because the P/2010 A2 dust particles have a composition that is similar to ordinary chondrites \citep{K12}, we assumed $\rho_\mathrm{d}$ = 3000 kg m$^{-3}$, which is a typical value for their bulk density \citep{Britt2002}. This assumption is consistent with previous research \citep{J10,J13,K13}. $K$ = 5.7 $\times$ 10$^{-4}$ kg m$^{-2}$ is a constant, and $Q_\mathrm{pr}$ is a radiation pressure coefficient that we considered to be unity \citep{Burns1979}. We supposed an impulsive dust ejection on 2009 March 2 following the previous research \citep{J10,J13,A13}. We employed a size-dependent terminal speed for the dust particles:

\begin{equation}
V_\mathrm{ej} = V_0 \left(\frac{\beta}{\beta_\mathrm{min}}\right)^{u_1}~,
 \label{eq:velocity}
\end{equation}

\noindent where $V_0$ is the reference ejection speed for the largest particles ($\beta=\beta_\mathrm{min}$). The exponent, $u_1$, is the power-index of the size-dependent ejection speed. The number of dust particles is given by

\begin{equation}
N(a_\mathrm{d})~da_\mathrm{d} = N_0 \left(\frac{a_\mathrm{d}}{a_{0}}\right)^{-q}~da_\mathrm{d}~, 
 \label{eq:size}
\end{equation}

\noindent in the size range of $a_\mathrm{min}$ $\leqslant$ $a_\mathrm{d}$ $\leqslant$ $a_\mathrm{max}$, where $a_\mathrm{min}$ and $a_\mathrm{max}$ are the minimum and maximum particle sizes given by $a_\mathrm{min}$=$KQ_\mathrm{pr}$/$\rho_\mathrm{d} \beta_\mathrm{max}$ and $a_\mathrm{max}$=$KQ_\mathrm{pr}$/$\rho_\mathrm{d} \beta_\mathrm{min}$, respectively. $q$ is the power-index of the differential size distribution. $N_0$ is the reference number of dust particles at the reference size of $a_0$=1~m.

\begin{figure}
\epsscale{0.8}
\plotone{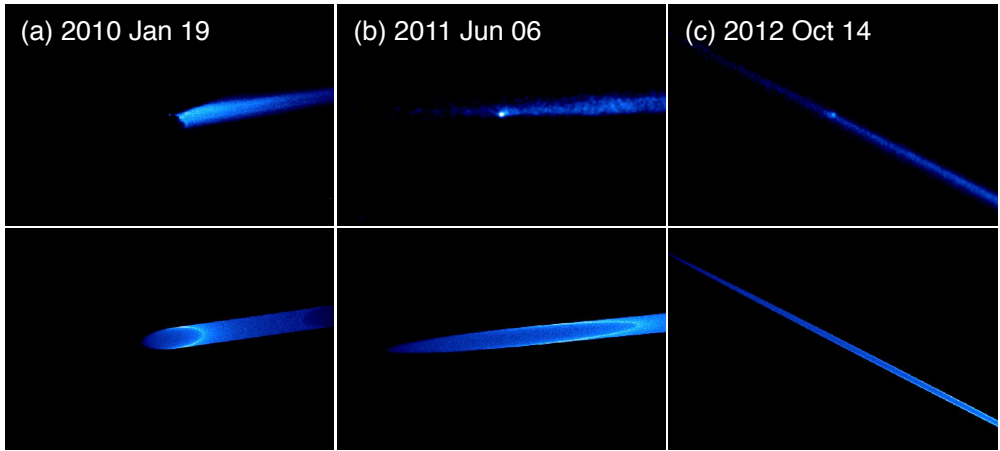}
\caption{Comparison between the observed images and the ``isotropic model" images at three different epochs. The top three figures are the observed images for (a) Gemini-2010, (b) Subaru-2011 and (c) Keck-2012, and the bottom three figures are the model images for (d) Gemini-2010, (e) Subaru-2011 and (f) Keck-2012. The parameters in these model images are the same as those in \citet{J13} (also see Table \ref{tab:parameter}), where they assumed that the DEP is located at the LF (i.e., the center of these images). All of these images have the FOV of $1.45\arcmin \times0.97\arcmin$ in the standard orientation: north is up and east is to the left. }
\label{fig:isotropic}
\end{figure}

The assumptions and model above are, in principle, the same as those in \citet{J13}, where they assumed a size-independent (i.e., $u_1$=0 in Equation (\ref{eq:velocity})) ``isotropic'' ejection (also see Table \ref{tab:parameter} for the model parameters of the isotropic model). Namely, they  considered that the dust particles were ejected in every direction without thinking of the size dependency of the ejection speed. In addition, \citet{J13} assumed that the dust particles were ejected from the observed position of the LF. Figure \ref{fig:isotropic} shows a comparison between the isotropic model and the observations at different epochs. It is true that the isotropic model reproduced the observed morphology and surface brightness distribution in the Keck-2012 image well, but in our opinion, it is not satisfactory to explain the time evolution of the dust cloud.  In particular, we would like to draw attention to the unique morphology observed in early 2010 (the arcs), which is not present in the isotropic ejection model images of the early 2010 observations. Therefore,  certain modifications are required to ensure consistency with the observations in early 2010.

\begin{figure}
\epsscale{0.75}
\plotone{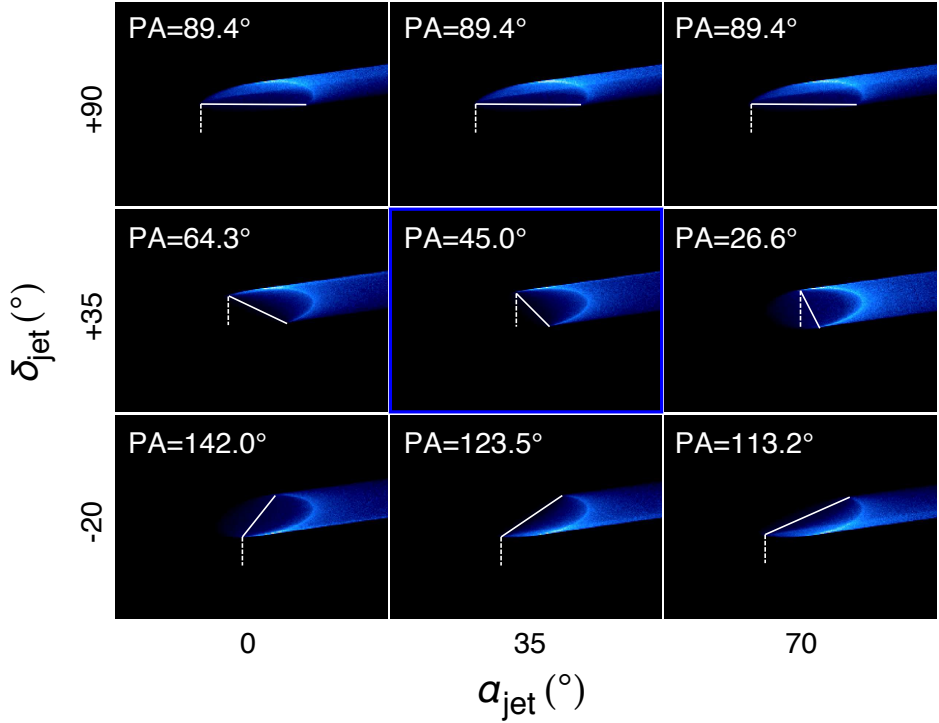}
\caption{Examples of the ``anisotropic model" images for the observation on 2010 January 19 considering a hemispherical (i.e., $w$=90\arcdeg) dust ejection with different jet orientations as labeled. Except $\alpha_\mathrm{jet}$, $\delta_\mathrm{jet}$, and $w$, we employed the same parameters as shown in Table \ref{tab:parameter}. In these figures, $\alpha_\mathrm{jet}$ values are constant along the column while $\delta_\mathrm{jet}$ values are constant along the row. In all panels, the DEP is located at the center.
To evaluate each images, we measured the position angle (PA) connecting two edges of the arc from the east edge to the west edge (solid lines), with respect to the south direction (dashed lines).}
\label{fig:search}
\end{figure}

There are several ideas for the creation of the arc-like features. For example, assuming little or  no radiation pressure, it would be possible to replicate  arc features that  are consistent with the images at a single epoch or for a certain short time duration \citep{A13,K13}. However, as we mentioned above, these models do not reproduce the morphology and brightness distribution at different epochs. We noticed that arc A can be produced without thinking of exquisite dust ejection models when we assumed a simple cone-shaped jet with a half opening angle of $w$ (not an impact hollow cone but a solid cone), although we initially did not consider the physical implication of the ejection model. Figure \ref{fig:search} shows the example of model images with a hemispherical (i.e., $w$=90\arcdeg) dust ejection considering different orientations of the central axis of the cone-shaped jet. It is clear that some model images show arc A at the eastern edge of the trail structure, which is similar to the observations in early 2010 (see Figure \ref{fig:obs} (a) and (b)). We noticed that the brightness enhancement of arc A can be explained by the high existence probability of the largest particles that were ejected at similar ejection velocities. Such particles tend to form a cut end of the dust trail at the leading edge when they were ejected toward the trailing direction of the orbital motion. For this reason, we modified the isotropic model to what we call an ``anisotropic model'', where we assumed that dust particles were ejected symmetrically with respect to a direction in the inertial frame (i.e., toward right ascension $\alpha_\mathrm{jet}$ and declination $\delta_\mathrm{jet}$) in a cone-shaped jet with a half opening angle of $w$, and searched for the best-fit parameter set to match the observed data. In addition, we left the position of the dust source (i.e., the dust ejection point, DEP) out of our consideration in order not to be fixated on the previous ideas (although the dust particles were assumed to have originated from the LF in the previous publications). In order to keep the constant width of the dust trail in the model, we imposed a constraint of the reference ejection speed of $V_0$ $\propto$ $\sin \left(w\right)^{-1}$ \citep[see, e.g.,][]{Ishiguro2016}. Once we obtained the positions of the dust particles in celestial coordinates by analytically solving the Kepler equation with solar radiation pressure, we calculated the cross-sectional areas of the dust particles in the CCD coordinate system:

\begin{equation}
C_\mathrm{pixel}(x,y) = \int^{a_\mathrm{max}}_{a_\mathrm{min}}N_\mathrm{cal}(a_\mathrm{d},x,y) \pi a_\mathrm{d}^2~da_\mathrm{d}~, 
\label{eq:modelflux}
\end{equation}

\noindent where $N_\mathrm{cal}(a_\mathrm{d},x,y)$ is the number of dust particles counted within a pixel at the coordinates $(x, y)$ in a CCD image. A full list of free parameters and the test range is shown in Table \ref{tab:parameter}.

\clearpage
\section{Results}
\label{sec:result}
\subsection{Dust Cloud Morphology and Surface Brightness}
\label{subsec:dustresult}
Among the eight unknown parameters in our model above ($\alpha_\mathrm{jet}$, $\delta_\mathrm{jet}$, $w$, $\beta_\mathrm{max}$, $\beta_\mathrm{min}$, $V_0$, $u_1$, $q$), we first determined the three parameters for a cone-shaped jet. 
We measured the position angle (PA) connecting two edges of arc A from the observation images of Gemini-2010, from the east edge to the west edge, PA=45.0$\arcdeg$ with respect to the south direction.
We then created a number of simulation images with a hemispherical (i.e., $w$=90\arcdeg) dust ejection considering different orientations of the central axis of the cone-shaped jet ($\alpha_\mathrm{jet}$, $\delta_\mathrm{jet}$) and measured the PA of each modeled arcs (Figure \ref{fig:search}). We found that the best-fit on arc A can be obtained when the particles ejected in a jet direction of 25$\arcdeg \la \alpha_\mathrm{jet} \la$ 40$\arcdeg$ and 30$\arcdeg \la \delta_\mathrm{jet} \la$ 40$\arcdeg$. The PA was independent of $w$.
Assuming a jet direction of $(35\arcdeg, 35\arcdeg)$,
$w$ was derived using the curvature $\kappa$ of the forth-order polynomials $f(x)$ fitted to the (x, y) positions of modeled arcs where we reversed x and y (i.e., (y, x)), at $x=x_0$ where it makes the most distant point from the line connecting two edges of the arc:
\begin{equation}
\label{eq:curvature}
\kappa = \frac{| f''(x) |}{\left[1+f'(x)^{2}\right]^{3/2}}~,
\end{equation}
\noindent where we measured $\kappa=4.07\times10^{-2}$ from the observation images of Gemini-2010.
We obtained 20$\arcdeg \la w \la$ 30$\arcdeg$.
Repeating the process using different jet directions in the plausible ranges gives the same solution of $w$.
Once we fixed ($\alpha_\mathrm{jet}, \delta_\mathrm{jet}, w$), we obtained the DEP as an outcome (see more descriptions in Section \ref{subsec:DEP}).

Second, we determined the smallest particles size ($\beta_\mathrm{max}$) and the power-index of the size-dependent ejection speed ($u_1$) using the Subaru-2011 image. Small particles were susceptible to the solar radiation pressure and difficult to  detect in images with narrow  fields of view (FOV). We found the Subaru-2011 image taken with the wide-field camera Suprime-Cam was the best to determine the smallest particle size because it has the largest orbital coverage (a delta mean anomaly of 0.42$\arcdeg$). We obtained $\beta_\mathrm{max}=7\times10^{-4}$ ($a_\mathrm{d}$=270 \micron) from the comparison of observation and model images to explain the existence of the dust trail that extended beyond the FOV of the Subaru-2011 image.

\begin{figure}
\epsscale{0.7}
\plotone{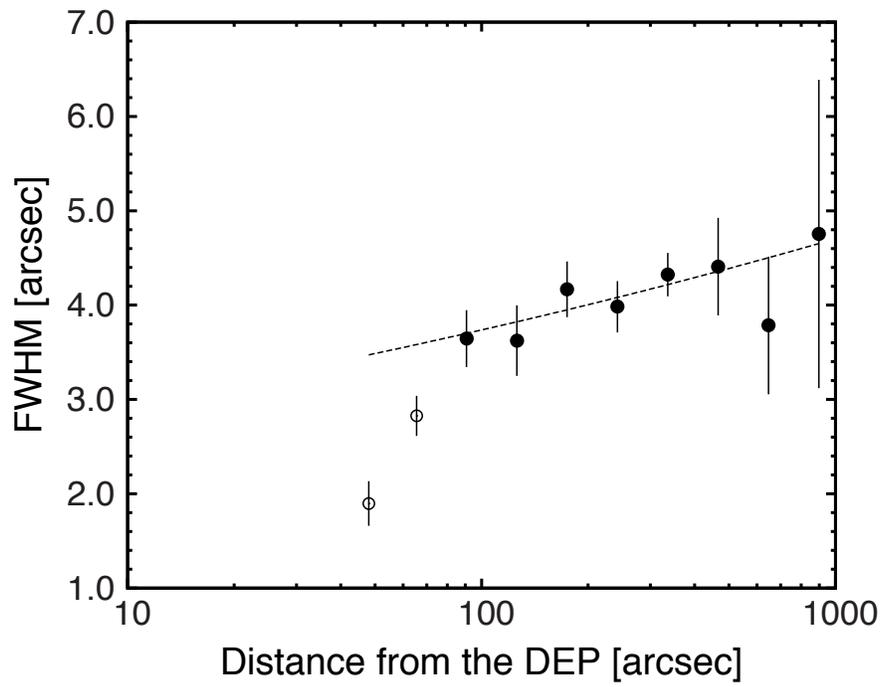}
\caption{The FWHM of the trail with respect to the distance from the LF measured using the Subaru-2011 image. The observed data beyond 80$\arcsec$ (filled circles) are fitted by an exponential function with an index of 0.10$\pm$0.02 (dashed line).
}
\label{fig:width}
\end{figure}

Figure \ref{fig:width} shows the width of the trail as a function of distance from the reference point (i.e., the DEP), where we measured the width from the FWHM of a series of surface cut profiles perpendicular to the trail. In the figure, we found that the trail widened sharply at 40--80\arcsec\ (open circles) and moderately beyond 80 \arcsec\ (filled circles) as the distance increased. We ignored the data at 40--80\arcsec\ because we might sampled the data from the fine structures (i.e.,  arc A and B) and fitted the slope with a power-law function (dashed line). Since smaller particles were distributed farther via the solar radiation pressure, the observed widening beyond 80 \arcsec\ suggests that the smaller particles were ejected with higher ejection velocity. We derived the power-index of the size-dependent ejection speed $u_1$=0.10$\pm$0.02 through the power-law fitting on the relation between width (i.e., ejection speed perpendicular to the orbital plane) and distance (i.e., proportional to $\beta$ assuming that the dust motion parallel to the orbit plane is determined by radiation pressure acceleration). Once we determined these five parameters ($\alpha_\mathrm{jet}$, $\delta_\mathrm{jet}$, $w$, $\beta_\mathrm{max}$ and $u_1$), we deduced plausible values of $V_0$, $\beta_\mathrm{min}$ and $q$, considering both morphology and surface brightness of the dust trail at multiple observation epochs. Through comparison with our dynamical model, we derived $V_0$ and $\beta_\mathrm{min}$ by fitting the overall trail width and the extent of arc A (i.e., cut end of the trail), which is sensitive to the ejection speed of the largest trail particles.
The best-fit $V_0$ and $u_1$ resulted in the maximum speed for the smallest particles of $\sim$0.50~m s$^{-1}$.

In Table \ref{tab:parameter}, we summarized the best-fit parameters for our dynamical model. The best-fit model shows good agreement with the observed morphology at any epochs in 2010--2012 (in the middle row of Figure \ref{fig:anisotropic}). Moreover, the model also fits the surface brightness distribution in a broad sense (Figure \ref{fig:sbr}), although there are modest differences near the peaks which can be improved by a fine tuning of the number of dust particles around the maximum size. Because our new model parameters were obtained by following the isotropic model, the results (especially the size and the size distribution) are consistent with those in \citet{J13}. However, the trivial modification to the anisotropic dust ejection resulted in a remarkable improvement in reproducing the observed morphology (the trail and arc A) at any observation epochs over $\sim$3 years. We would like to insist that this is the first success, which demonstrates the consistency with both the time-variant morphology and the surface brightness profiles simultaneously. The remaining features which is not produced in our model is arc B, which will be discussed later.

\subsection{The Location of the LF and the Dust Ejection Point (DEP)}
\label{subsec:DEP}
\begin{figure}
\epsscale{0.9}
\plotone{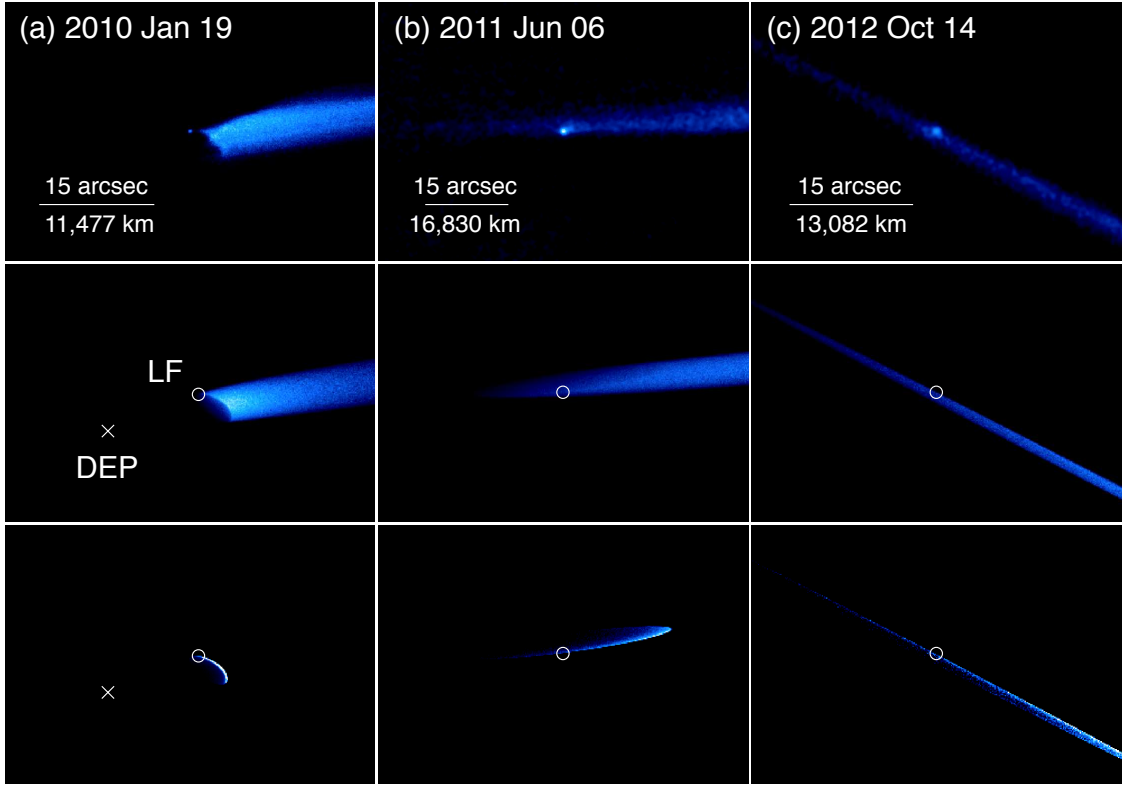}
\caption{The observation images at three different epoch (top), the anisotropic model images with the best-fit parameters (middle) and the existence probability maps of the large fragments (bottom) for the observation of Gemini-2010 (a, d, g), Subaru-2011 (b, e, h) and Keck-2012 (c, f, i). The positions of the LF and DEP are indicated by open circles and crosses, respectively. We do not show the positions of the DEP in the second and third columns (Subaru-2011 and Keck-2012) because they exist beyond the FOV. The best-fit model parameters are summarized in Table \ref{tab:parameter}.
}
\label{fig:anisotropic}
\end{figure}

\begin{figure}
\epsscale{0.7}
\plotone{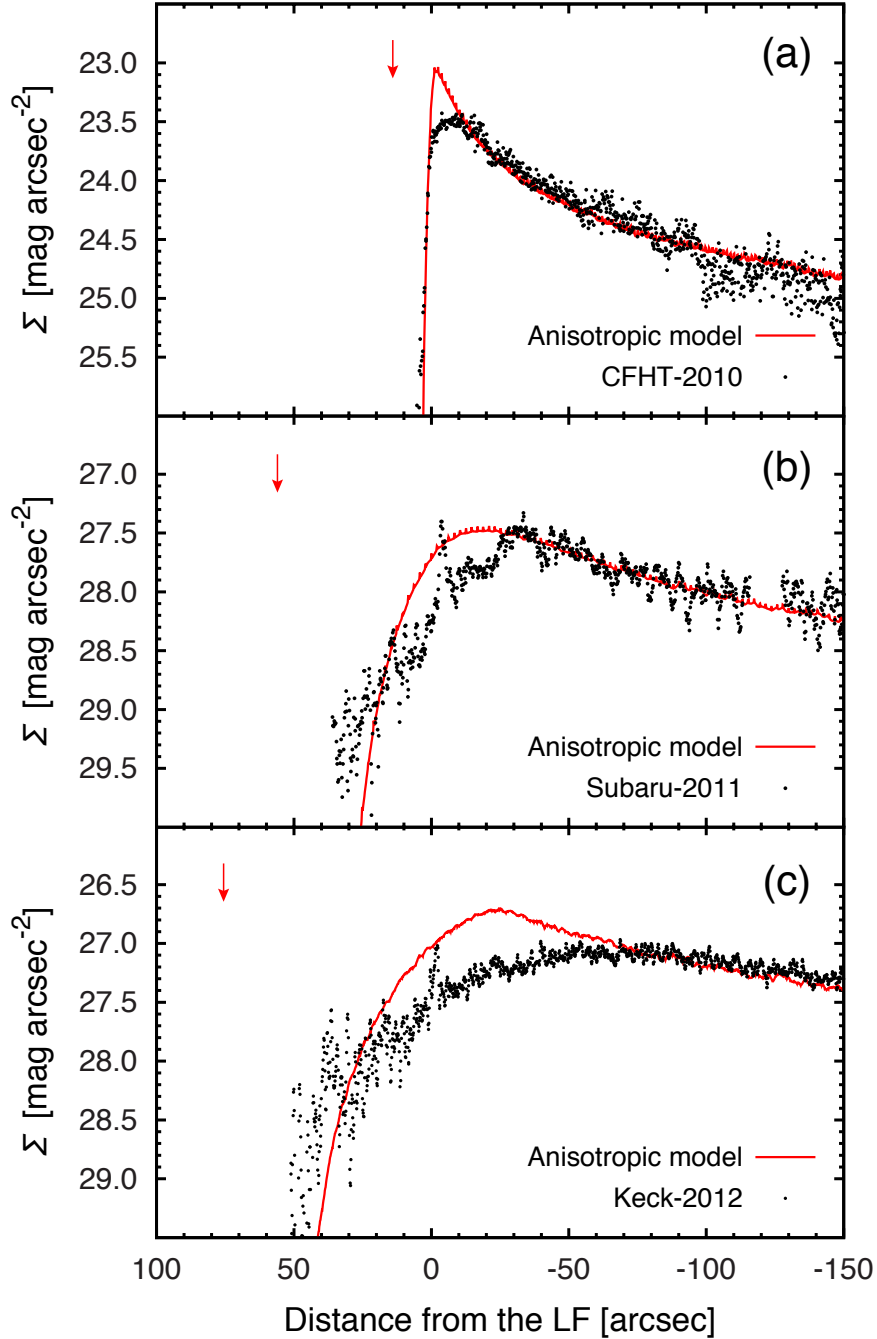}
\caption{Comparison of the surface brightnesses ($\Sigma$) between the observations and the best-fit anisotropic model. These values were measured along the axis of the dust trail for the observations of (a) CFHT-2010 ($R$-band), (b) Subaru-2011 ({\sl g}$'$-band) and (c) Keck-2012 ($B$-band). Distances are measured with respect to the position of LF. The locations of the DEPs in the model are indicated by arrows. 
}
\label{fig:sbr}
\end{figure}

As we mentioned in Section \ref{sec:model}, we did not specify any ``detectable  objects'' as the dust source and determined the model parameters in Section \ref{subsec:dustresult}. Since the LF had been considered to be the dust source in all previous research, it is important to examine the location of the LF with respect to the dust cloud in our model simulation. We hence calculated the positions of $\approx$10$^6$ test particles using the same anisotropic model, where we fixed $\beta=0$ and $V_\mathrm{ej}=V_0$ to take into account only the large particles without the solar radiation pressure effect. We found that the big test particles ($\beta=0$) tend to appear along the arc with a high probability, where the arc is morphologically identical to arc A ($\beta=\beta_\mathrm{min}$) with a  negligible offset of $\sim$0.4\arcsec~(i.e., unresolved under the ground-based seeing disk size), because such large dust particles ($\beta\le10^{-6}$) are almost stationary against the solar radiation pressure. Figure \ref{fig:anisotropic} (bottom row) shows the existence probability of the LF together with the observation images (top) and dust model images (middle) for three different epochs in 2010, 2011 and 2012. Over the years, the positions of the largest test particles show good agreement with the modeled arc A. From the result, we conjectured that intermediate sized fragments (from decimeter to hundred-meter particles) might distribute along arc A with high probability but only the LF had been detected because it is brighter than the detection limits of these observations. In Figure \ref{fig:anisotropic}, crosses show the location of the DEP.
It is important to note that the DEP deviated from the observed position of the LF. This result implies that dust particles were not ejected from the LF, but from a position where no object was detected from any observations (which will be discussed later). Similar trends can be seen in the surface brightness profiles (Figure \ref{fig:sbr}). Although the derived parameters are very similar between the isotropic model and our anisotropic model, these two differ in that the former assumed that the dust ejection was from the LF, while the latter assumed that it was from a position where no object had been detected.

We examined the orbital difference between the LF and the DEP. We analyzed the positions of the LF and the DEP in $\sim$40 epoch observations with relevant model images and derived the orbital elements in the J2000 coordinate system using the Find_Orb\footnote{http://www.projectpluto.com/find_orb.htm} software package. Table \ref{tab:orbit} shows the osculating orbital elements ($a$,$e$,$i$) of the LF and the DEP. We found the best set of orbital elements with a negligible residual in the celestial plane ($\sim$0.7\arcsec). The orbital elements of the DEP is in consistent with those of the LF down to $\sim$4th decimal place but significantly different from them to an accuracy of the uncertainties (around $\sim$5th decimal place, from NASA/JPL Small-Body Database Browser\footnote{http://ssd.jpl.nasa.gov/}). To confirm if the orbital elements of the LF are available for our dust model, we performed another set of anisotropic model simulations using the new orbital elements for the DEP, but we could not find any notable differences in the modeled dust morphology. This means that the trivial difference in the orbital elements does not change the above results for our dust model simulation.

\subsection{Motions of the Fragments}
\label{subsec:fragment}

\begin{figure}
\epsscale{0.55}
\plotone{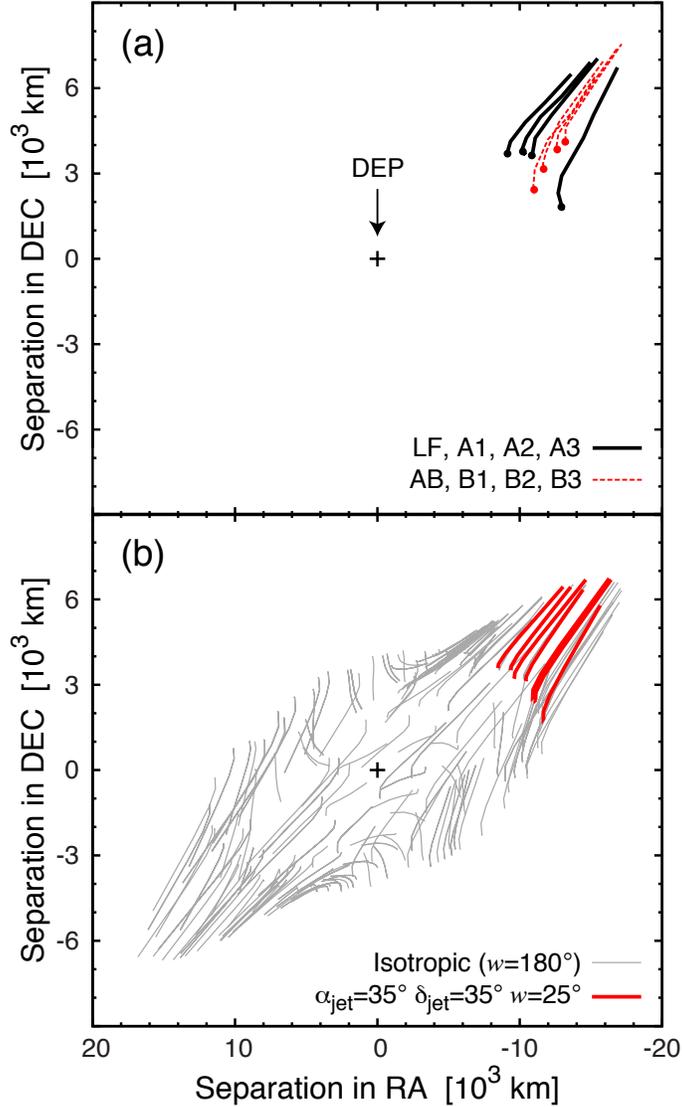}
\caption{The motions of the fragments in a series of HST-2010 images from UT 2010 January 25 to UT 2010 May 8. (a) shows the observed motion of eight fragments with respect to the DEP. The positions at the first observations on UT 2010 Jan 25 are marked with filled circles. (b) displays the simulated trajectories of large ($\beta=0$) fragments ejected in every direction with $V_\mathrm{ej}$=0.28~m s$^{-1}$ (thin lines). Thick red lines show trajectories of seven sampled test particles ejected within a cone whose central axis points in the direction of $(35\arcdeg, 35\arcdeg)$ having a half-opening angle of $w=25\arcdeg$.
}
\label{fig:fragmotion}
\end{figure}

We performed further analysis on the motions of the fragments distributed along arcs A and B, following the designation in \citet{A13}. The motions of the fragments were thoroughly studied in \citet{A13}, where they regarded the LF as the source and investigated the motion with respect to the LF. However, because the fragments and dust particles are not ejected from the LF in our above analysis, we should reconsider the motion with respect to the DEP we derived above. Figure \ref{fig:fragmotion} (a) shows the observed  sky-plane trajectories of fragments from UT 2010 January 25 to May 8 with respect to the DEP, showing the positions at the first epoch of the HST-2010 image as the filled circles. We show the motions of not only these fragments (labeled with A1, A2, A3, AB, B1, B2 and B3) but also the LF because the LF is no longer the source of the materials. We found that all fragments were moving toward the northwest in the observed frame.

To give an interpretation to the motion, we examined the trajectories of the fragments (i.e., $\beta$=0) through a dynamical analysis. Figure \ref{fig:fragmotion} (b) shows the calculated trajectories of the fragments ejected from the DEP on UT 2009 March 2 (the same day as the dust ejection) with different ejection directions. We employed the orbital elements of the DEP derived in Section \ref{subsec:DEP} and considered the fragment ejection in every direction (i.e., isotropic ejection) to think about all possibilities. We found that the observed trajectories can be explained only when these objects were ejected in the same direction as the dust particles in our anisotropic dust model (Table \ref{tab:parameter}). The agreement suggests that the large fragments were ejected together with the small dust particles from the DEP in the same direction (i.e., 25$\arcdeg \la \alpha_\mathrm{jet} \la$ 40$\arcdeg$ and 30$\arcdeg \la \delta_\mathrm{jet} \la$ 40$\arcdeg$ with 20$\arcdeg \la w \la$ 30$\arcdeg$). We derived the typical ejection speed of the fragments of $V_\mathrm{ej}$=0.28~m s$^{-1}$ from this dynamical analysis, which is consistent with the best-fit $V_0$ (the velocity of the largest dust particles) value in our anisotropic dust model.

\begin{figure}
\epsscale{0.35}
\plotone{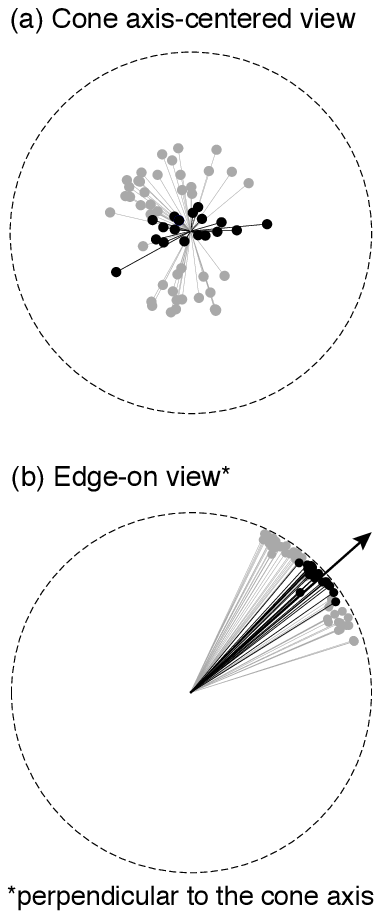}
\caption{The projected ejection velocity of particles in group A (gray) and group B (black).  (a) is the cone axis-centered view, where the jet emerges in the direction perpendicular to this figure, while (b) is the edge-on view, where the jet direction of $(35\arcdeg, 35\arcdeg)$ is indicated by the arrow.
Filled circles denote the points on the sphere, while sold lines are the unit velocity vectors of each particle. The cross-sectional velocities on this figure distribute on the dashed lines.
}
\label{fig:initial}
\end{figure}

In Figure \ref{fig:fragmotion} (a), the trajectories of the fragments on arc B (i.e., AB, B1, B2 and B3) concentrated on the narrow region and aligned parallel to the bulk motion of the fragments projected onto the sky plane. To explain the trend on arc B, we further ran a dynamical simulation of 100 test particles using our best-fit fragment model (Table \ref{tab:parameter})
and compared them to the observed trajectories of the eight fragments. For convenience of classification, we divided them into two groups: group A for the fragments having similar trajectories to the LF, A1, A2 and A3 (i.e., arc A) and group B for the fragments having similar trajectories to B1, B2 and B3 (i.e., arc B).
In the case of AB-like fragments that intersect both arcs A and B, we regarded them as group A.
By visual inspection, we classified 100 test particles into two groups, that is, 53 and 19 particles are in groups A and B, respectively. Since we strictly selected the test particles only when they are almost identical to those fragments, 28 test particles were not classified into any group.
We then recorded the initial velocity information about the selected particles in equatorial rectangular coordinates, $(v_x,v_y,v_z)$, and reconstructed the ejection velocity field at the moment of disruption as shown in Figure \ref{fig:initial}.
From the cone axis-centered view (Figure \ref{fig:initial} (a)), we found that group B particles (black) have a limited spatial distribution, while group A particles (gray) have an almost isotropic distribution within a cone-shaped jet.
The edge-on view (Figure \ref{fig:initial} (b)) suggests that the ejection velocity field of group B particles is almost parallel to the jet direction (i.e., the bulk motion of fragments), as we expected from Figure \ref{fig:fragmotion} (a). For quantitative analysis, we can consider the total unit velocity vector with respect to the DEP, $\hat{\mathbf{v}}_\mathrm{tot}$,

\begin{eqnarray}
\begin{array}{l}
\displaystyle \mathbf{v}_\mathrm{tot} = (v_{x,\mathrm{tot}}, v_{y,\mathrm{tot}}, v_{z,\mathrm{tot}}) = \sum\limits_{i}^\mathrm{Ntp} (v_{x,i}, v_{y,i}, v_{z,i})~,\\
\displaystyle \hat{\mathbf{v}}_\mathrm{tot} = \frac{\mathbf{v}_\mathrm{tot}}{(v_{x,\mathrm{tot}}^{2} + v_{y,\mathrm{tot}}^{2} + v_{z,\mathrm{tot}} ^{2})^{1/2}}~,
\end{array} 
\label{eq:vector}
\end{eqnarray}

\noindent where we calculated the total unit velocity vectors as $\hat{\mathbf{v}}_\mathrm{tot,A}=(0.717,0.389,0.579)$ and $\hat{\mathbf{v}}_\mathrm{tot,B}=(0.672,0.457,0.583)$ for the group A and B fragments, respectively.
It is notable that $\hat{\mathbf{v}}_\mathrm{tot,B}$ has a negligible separation of $\sim$0.9$\arcdeg$ with respect to the jet centroid $(35\arcdeg, 35\arcdeg)$, while $\hat{\mathbf{v}}_\mathrm{tot,A}$ has a relatively wide separation of $\sim$5.3$\arcdeg$.
Further discussion will be provided in Section \ref{subsec:impact}.

\clearpage
\section{Discussion}
\label{sec:discussion}

For both dust particles and fragments, key features to be explained include

\noindent 1. Absence of the central body at the DEP;\\
2. Anisotropic ejection with a limited angle; \\
3. Low ejection speed ($\ll$ 1~m s$^{-1}$); \\
4. Similarity in ejection speeds between fragments and  largest dust particles ($V_\mathrm{ej} \approx V_0$).

\subsection{Ejection mechanism}
\label{subsec:mechanism}

Four ejection mechanisms have been suggested so far, that is, sublimation of ice \citep{M10}, rotational breakup \citep{J10,Marzari2011,A13}, impact cratering \citep{H12,K13} and impact shattering \citep[also called ``catastrophic disruption'';][]{J10,S10,K12}. Here, we evaluate four mechanisms with two fundamental questions: (i) Does it require momentum conservation on the original body? (ii) Does it require the central body (i.e., the LF) survived at the DEP? 

To answer the first question (i), sublimation and rotational breakup require momentum conservation on the original body before and after the mass loss (i.e., total ejecta momentum should be zero with respect to the DEP), where we assumed there is no external force. On the other hand, we expect non-conservation of momentum from an impact unless we take into account the external momentum because impact projectile injects momentum into the target asteroid.
In the case of P/2010 A2, most of the mass is occupied by large fragments rather than small dust particles, while the ejection velocities are approximately the same regardless of the size, suggesting that the largest bodies make up a significant proportion of the total momentum. We thus consider the sum of momentum of the eight largest fragments with respect to the DEP as a proxy for the total momentum in the system and found that it never converges to zero in our model (cf. Section \ref{subsec:fragment}). Independently, we revisited a rotational breakup model where the DEP was assumed to be the LF \citep{A13} and considered the total momentum of seven sub-fragments with respect to the LF. All sub-fragments had negative velocities in the y- and z-directions, meaning that the total momentum cannot converge to zero regardless of the individual fragment masses and x-velocities. We conclude that momentum is not conserved on the original body of P/2010 A2, which rules out the two mechanisms requiring zero total momentum, i.e., sublimation and rotational breakup.

The second question (ii) is actually a criterion to judge the degree of the fragmentation of the target asteroid to differentiate the two remaining possibilities, i.e., impact cratering and shattering. Both are impacts, but a shattering (i.e., complete target destruction) require more ``specific energy'' (impact kinetic energy per total mass of the system) than a cratering (i.e., partial target destruction). The central body must survive from an impact cratering, whereas it may not survive from an impact shattering, leaving nothing at the DEP but produce interplanetary debris.
The point of the issue is summarized in Table \ref{tab:mechanism}. From the two aspects that (i) there is a non-zero value of the total momentum in the system and (ii) no central body existed at the DEP, we arrived at the conclusion that the ejection mechanism of P/2010 A2 is an impact shattering.

\subsection{Impact shattering interpretation}
\label{subsec:impact}
\subsubsection{Comparison to laboratory experiments}

\begin{figure}
\epsscale{0.7}
\plotone{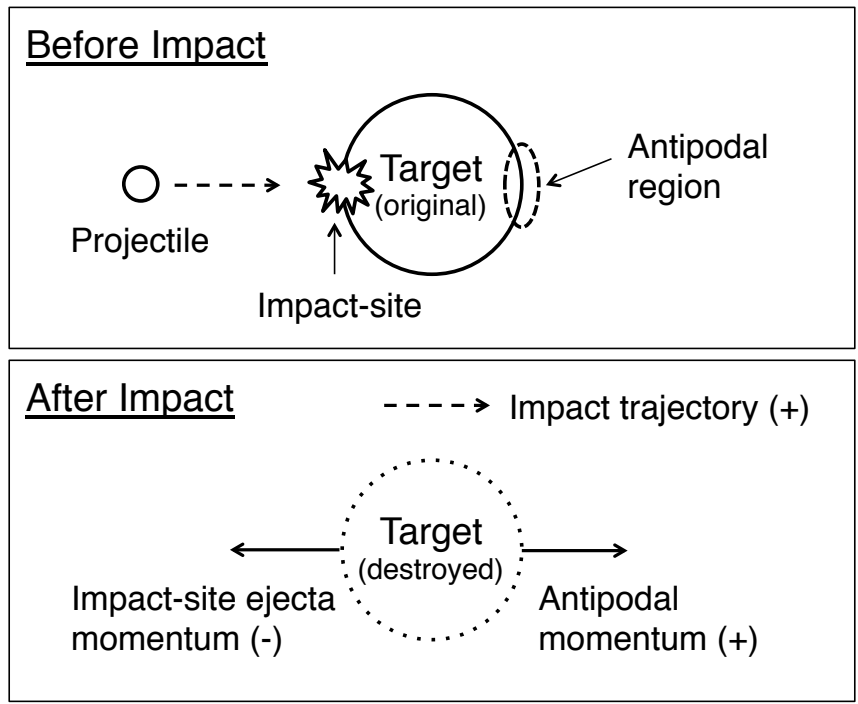}
\caption{A conjectured configuration of the impact before (top) and after (bottom) the event. The components of the impact-site ejecta momentum $(-)$, the antipodal momentum $(+)$ and the impact trajectory $(+)$ are illustrated by arrows. In this scenario, we conjecture that small and fast particles would be produced at the impact-site and move in the opposite direction of the impact trajectory, while the largest and slowest fragments could be generated in the antipodal region (Section \ref{subsec:impact}, Equation \ref{eq:impact1}) and move in the direction parallel to the impactor.
}
\label{fig:antipodal}
\end{figure}

To verify our hypothesis that P/2010 A2 is the debris from an impact shattering, we compare the results of our simulation to those of the laboratory impact experiments from the literature.
In the laboratory experiments using various targets, it is commonly observed that small ($\la100~\micron$) particles with high velocities ($>$10~m s$^{-1}$) are produced at the point of impact in the opposite direction of the impact trajectory (cf. Figure \ref{fig:antipodal}), while the largest and slowest fragments are usually located directly opposite the impact site, which we call ``antipodal'' fragments \citep{Asada1985,N91}. Specifically for targets of basalt and gypsum, the antipodal region suffers the least fragmentation, and a number of large fragments are generated with a limited distribution around the antipodal point; it has been observed that such fragments have similar velocities \citep{Giblin1998}.
Considering the typical antipodal velocity of $\sim$(5--10)~m s$^{-1}$ in the shattering experiments,
we notice that there is a velocity discrepancy between P/2010 A2 ($\ll$ 1~m s$^{-1}$) and the laboratory counterparts. To eliminate the discrepancy and explain the low antipodal velocities, we consider two approaches as follows.

The first approach is making the target weaker so that it is  easily shattered by a relatively small specific energy, i.e., adopting the idea by \citet{Fujiwara1987} that antipodal velocities decreased with decreasing specific energy.
The key for low antipodal velocities will be to decrease the specific energy to as small as possible until it is enough to shatter a ``weak'' body.
It has been suggested that a target with a low static strength should be easier to  shatter via impact than high strength targets \citep{DR90}, while it is also known that sub-kilometer to kilometer sized bodies are significantly weaker than bodies in other sizes \citep{HH99,Jutzi2010,J10,J13}. 
Once the specific energy is enough to destroy a body, the second approach is damping the propagation of the shock wave through the target, sheltering the antipodal region from significant damage while intensifying the deposition of the shock energy at the impact point \citep{Asphaug1998}. It was reported that porous materials (e.g., gypsum and sandbags) have lower antipodal velocities than those of nonporous materials (e.g., basalt and ice) because  pre-existing fractures and voids in the target body cause a rapid attenuation of shock pressure \citep{YI94,OA2009}.

To summarize the comparison between the results from P/2010 A2 and the laboratory experiments, we found that the large fragments generated around the antipodal point provide the best match to our results, i.e., anisotropic ejection with a limited angle and almost constant ejection speeds for large ejecta ($V_\mathrm{ej} \approx V_0$). To explain the velocity discrepancy, we considered two possible approaches to reduce the antipodal velocities: (i) making the target weaker so that it is easily shattered by a relatively small specific energy (e.g., size and strength) and (ii) damping the propagation of the shock wave through the target (e.g., porosity). 
Combining all of the possible material properties of the affected asteroid, we speculate that the original body of P/2010 A2 could be a sub-kilometer sized rubble-pile asteroid like (25143) Itokawa (i.e., a porous and low static strength asteroid).

\subsubsection{Energy estimation}

Given the above considerations, to determine how realistic it is that the target asteroid can be shattered and produce low antipodal velocities,
we can estimate the specific energy delivered to the target asteroid.
The momentum conservation \textit{taking into account} the injected projectile momentum can be written as (also see the configuration in Figure \ref{fig:antipodal})

\begin{equation}
\label{eq:impact1}
p_\mathrm{proj} \hat{\mathbf{e}}_\mathrm{p} = -p_\mathrm{ej} \hat{\mathbf{e}}_\mathrm{p} + \Delta p_\mathrm{target} \hat{\mathbf{e}}_\mathrm{p}~,
\end{equation}

\noindent where $\hat{\mathbf{e}}_\mathrm{p}$ is the unit vector of the direction along the projectile momentum, $p_\mathrm{proj}$ and $p_\mathrm{ej}$ are the components of the projectile and escaping ejecta momentum, respectively. A minus sign appeared in the ejecta term because the majority of the ejecta are expected to be generated in the opposite direction of the impact trajectory \citep{N91}. $\Delta p_\mathrm{target}$ denotes the resulting momentum of the target along the impact direction (i.e., antipodal component).
Assuming $\Delta p_\mathrm{target} \sim M_\mathrm{anti} v_\mathrm{anti}$ where $M_\mathrm{anti}$ and $v_\mathrm{anti}$ are the sum of the antipodal fragment masses and the mean antipodal velocity (i.e., 0.28~m s$^{-1}$), respectively, we can rewrite Equation \ref{eq:impact1} as

\begin{equation}
\label{eq:impact2}
m_\mathrm{p}V_\mathrm{impact} + (\mathrm{ejecta~momentum}) = M_\mathrm{anti} v_\mathrm{anti}~,
\end{equation}

\noindent where $m_\mathrm{p}$ and $V_\mathrm{impact}$ are mass of the projectile and the impact velocity, respectively. 
We assume $V_\mathrm{impact} \sim$ 5~km s$^{-1}$, which is the average collision velocity in the main asteroid belt \citep{5km}.
This indicates that

\begin{equation}
\label{eq:impact3}
M_\mathrm{anti} v_\mathrm{anti} \ga 2 \times m_\mathrm{p}V_\mathrm{impact}~,
\end{equation}

\noindent for collision with velocities higher than 5~km s$^{-1}$ \citep{HH12,HH15}. In this case, the specific energy (impact kinetic energy per total mass of the system) is given as

\begin{equation}
\label{eq:impact4}
Q^{\ast} \sim \frac{1}{2}\frac{m_\mathrm{p}}{M_\mathrm{anti}}V_\mathrm{impact}^2~,
\end{equation}

\noindent where we assume that the total mass of the system (i.e., the sum of projectile and target mass) can be approximated by $M_\mathrm{anti}$.
Combining Equation \ref{eq:impact3} and \ref{eq:impact4} gives

\begin{equation}
\label{eq:impact5}
Q^{\ast} \la~\frac{1}{2}~\frac{v_\mathrm{anti}} {2 V_\mathrm{impact}} V_\mathrm{impact}^2 = \frac{1}{4} v_\mathrm{anti} V_\mathrm{impact}~,
\end{equation}

\noindent and we obtained $Q^{\ast} \la$ 350~J kg$^{-1}$.
Interestingly, this energy density corresponds closely to the shattering threshold, $Q^{\ast}_{S}$, of a ten-meter diameter body in a recent laboratory examination using porous gypsum targets \citep[$Q^{\ast}_{S} \sim$300~J kg$^{-1}$ for 3~km s$^{-1}$ normal collision;][]{Nakamura2015}, where they are also in good agreement with that of tens to hundreds of meters bodies estimated by the numerical simulation \citep[$Q^{\ast}_{S} \sim$200--400~J kg$^{-1}$ for 3~km s$^{-1}$ and 45 degree oblique collision;][]{Jutzi2010}.
We thus conjecture that such small values of $Q^{\ast}$ delivered to the original asteroid enable an impact shattering ($Q^{\ast} \ga Q^{\ast}_{S}$), resulting in low antipodal velocities down to $\ll$1~m s$^{-1}$.

\subsubsection{Remaining morphological interpretation}

Further details should be left as open questions. However, here, we suggest a possible scenario that all observed ejecta (LF+A+B+dust) originated from the antipodal region in the target asteroid and ejected with a similar ejection speed of $\sim$0.28~m s$^{-1}$. Of the ejecta, three components (the LF, arc A, and dust trail) essentially constructed arc A and connected the dust trail. On the contrary, the distinct morphology of arc B remains unexplained so far,
implying that it was created by different mechanism. We note that the simulated ejection velocity field of the B fragments was parallel to the jet direction (Section \ref{subsec:fragment}) and suggest that the injected momentum from the projectile to the target asteroid created a number of antipodal fragments along the direction of the momentum transfer (i.e., the impact trajectory). On the other hand, there should be more ejecta that originated from somewhere in the target asteroid but not from the antipodal region (cf. Equation \ref{eq:impact1}). Most of those particles are small and fast, as they underwent greater impact fragmentation, and they should be pushed out of the FOV by radiation pressure during the $\sim$1 year between the disruption in 2009 and the first discovery in 2010. Some particles with intermediate sizes and velocities may have a chance to remain in the FOV at the time of early observations, as we can see from the outer diffuse sources (cf. Gemini-2010 and HST-2010) but not from Subaru-2011 and Keck-2012.

We remark that above scenario would be one of the possible interpretations that ensure consistency with the observational evidence. The ratio of antipodal to non-antipodal debris is highly dependent upon target property or the distance between the impact point and the antipodal point, and the majority of slow debris from impact shattering does not necessarily have to be antipodal. The work presented here suggests a possibility of large amounts of antipodal debris existed as we see from \citet{Setoh2010}.

\clearpage
\section{Summary and Conclusions}
\label{sec:summary}

In this paper, we performed dust model simulations assuming an anisotropic ejection from P/2010 A2 and succeeded in reproducing the time-variant features in the archival observations over $\sim$3 years from 2010 January to 2012 October.
When we assumed that the dust particles and fragments were ejected in the same direction from a point (i.e., the dust ejection point, DEP) where no object had been detected in any observations, our anisotropic model can explain all of the observations including (i) the unique dust cloud morphology, (ii) the trail surface brightness and (iii) the motions of the fragments. Our major finding is that the DEP is decoupled from the largest fragment (cf. LF$=$DEP had been considered in all previous research).

Comparing our results to the laboratory impact experiments, we speculated about the regional variation in the degree of fragmentation, the ejection velocity field, the specific energy, and the physical properties such as the size, porosity and static strength of the impacted asteroid:

\begin{itemize}
\item The least fragmentation around the antipodal point of the shattered asteroid is comparable to an anisotropic ejection in the limited ejection velocity field.
\item The asteroid underwent an impact with the specific energy of $Q^{\ast} \la$ 350~J kg$^{-1}$.
\item Impacts on sub-kilometer sized rubble-pile asteroids like (25143) Itokawa may produce low antipodal ejection velocities down to $\ll$1~m s$^{-1}$.
\end{itemize} 

Finally, we remark that our results based on observations and their modelings are consistent with those obtained through laboratory impact experiments. The consistency supports the idea that the P/2010 A2 event is the first evidence of the impact shattering (i.e., total disruption) occurred in the present main asteroid belt.

\acknowledgments
This work was supported by two research programs through the National Research Foundation of Korea (NRF) funded by the Korean government (MEST) (No. 2012R1A4A1028713, 2015R1D1A1A01060025). Yoonyoung Kim was supported by the Kwanjeong educational foundation scholarship and the Fellowship for Fundamental Academic Fields. Junhan Kim and Hidekazu Hanayama helped with the Subaru observation (Program S11A-038). Marco Micheli gave comments on orbital determination. We also thank to Prof. Masahiko Arakawa for a fruitful discussion. This research utilized the Keck Observatory Archive (KOA) and the facilities of the Canadian Astronomy Data Centre. We also thank anonymous referee for careful reading of the manuscript.

\clearpage
\begin{table}
\footnotesize
\tablenum{1}
\caption{Observation summary}
  \begin{center}
    \begin{tabular}{lcccccccc}
\hline
UT date & Telescope & Filter & R$_\mathrm{h}$\tablenotemark{a} & $\Delta$\tablenotemark{b} & $\alpha$\tablenotemark{c} & $\nu$\tablenotemark{d} & Abbreviation\tablenotemark{e} & References\tablenotemark{f} \\
\hline
2009 Mar 02 & (Possible date of disruption) & -- & 2.269 & 3.224 & 5.6 & -93 & -- & -- \\
2010 Jan 19 & Gemini & {\sl r}$'$ & 2.015 & 1.055 & 8.5 & 17 & Gemini-2010 & [2] \\
2010 Jan 25 & HST & F606W & 2.018 & 1.078 & 11.5 & 19 & HST-2010 & [1] \\
2010 Jan 29 & HST & F606W & 2.019 & 1.099 & 13.5 & 20 & HST-2010 & [1] \\
2010 Feb 09 & CFHT & $R$ & 2.026 & 1.175 & 18.7 & 25 & CFHT-2010 & -- \\
2010 Feb 22 & HST & F606W & 2.034 & 1.286 & 23.1 & 29 & HST-2010 & [1] \\
2010 Mar 12 & HST & F606W & 2.047 & 1.473 & 27.0 & 36 & HST-2010 & [1] \\
2010 Apr 02 & HST & F606W & 2.066 & 1.717 & 28.8 & 43 & HST-2010 & [1] \\
2010 Apr 19 & HST & F606W & 2.083 & 1.922 & 28.7 & 49 & HST-2010 & [1] \\
2010 May 08 & HST & F606W & 2.105 & 2.150 & 27.4 & 55 & HST-2010 & [1] \\
2010 May 29 & HST & F606W & 2.130 & 2.393 & 25.0 & 62 & HST-2010 & [1] \\
2011 Jun 06 & Subaru & {\sl g}$'$ & 2.556 & 1.547 & 3.2 & 161 & Subaru-2011 & -- \\
2012 Oct 14 & Keck & $B$ & 2.189 & 1.202 & 5.2 & -76 & Keck-2012 & [3] \\
\hline
    \end{tabular}
  \end{center}
\tablenotetext{a}{Heliocentric distance (au)}
\tablenotetext{b}{Geocentric distance (au)}
\tablenotetext{c}{Solar phase angle (degrees)}
\tablenotetext{d}{True anomaly (degrees)}
\tablenotetext{e}{Abbreviation to refer the observation in the text}
\tablenotetext{f}{References: [1] \citet{J10}, [2] \citet{H12}, [3] \citet{J13}}
\label{tab:observation}
\end{table}

\clearpage
\begin{table}
\footnotesize
\tablenum{2}
  \caption{Model parameters}
  \begin{center}
    \begin{tabular}{llccc}
\hline
Parameter   & Input Values & Best-fit Values & Best-fit Values & Best-fit Values \\
 & (for anisotropic model) & (Isotropic)\tablenotemark{1} & (Anisotropic)\tablenotemark{2} & (Fragment)\tablenotemark{3} \\
\hline
$u_1$ & 0.1 (fixed) & 0 & 0.1 & --\\
$q$ & 3.0--4.0 with 0.1 interval & 3.5 & 3.5 & -- \\
$\beta_\mathrm{max}$ & 7$\times$10$^{-5}$--1$\times$10$^{-3}$ with 1$\times$10$^{-5}$ interval & 7$\times$10$^{-5}$ & 7$\times$10$^{-4}$ & 0 \\
$\beta_\mathrm{min}$ & 1$\times$10$^{-7}$--9$\times$10$^{-6}$ with 1$\times$10$^{-6}$ interval & 1$\times$10$^{-6}$ & 1$\times$10$^{-6}$ & 0 \\
$V_0$ (m s$^{-1}$) & 0.03--0.53 with 0.05 interval & 0.15 & 0.28 & 0.28 \\
$w$ (deg) & 15--90 with 5 interval & 180 & 20--30 & 20--30 \\
$\alpha_\mathrm{jet}$ (deg) & 0--360 with 5 interval & -- & 25--40 & 25--40 \\
$\delta_\mathrm{jet}$ (deg) & -90 to +90 with 5 interval & -- & 30--40 & 30--40 \\
\hline
    \end{tabular}
  \end{center}
 \tablenotetext{1}{\citet{J13}}
  \tablenotetext{2}{This work ($\beta \neq 0$ for dust particles)}
  \tablenotetext{3}{This work (fragment; $\beta = 0$ for fragments)}
 \label{tab:parameter}
\end{table}

\clearpage
\begin{table}
\tablenum{3}
  \caption{Orbital elements of the largest fragment (LF) and the dust ejection point (DEP)}
  \begin{center}
    \begin{tabular}{lccc}
\hline
   & $a$\tablenotemark{a} & $e$\tablenotemark{b} & $i$\tablenotemark{c} \\
\hline
P/2010 A2 (LF) & 2.29008 & 0.12479 & 5.25389 \\
P/2010 A2 (DEP) & 2.29005 & 0.12480 & 5.25353 \\
\hline
    \end{tabular}
  \end{center}
\tablenotetext{a}{Osculating semimajor axis (au)}
\tablenotetext{b}{Osculating eccentricity}
\tablenotetext{c}{Osculating inclination (degrees)}
 \label{tab:orbit}
\end{table}

\clearpage
\begin{table}
\tablenum{4}
  \caption{Mechanism evaluation chart}
  \begin{center}
    \begin{tabular}{lcc}
\hline
Mechanism & Momentum cons.\tablenotemark{1} & Central body\tablenotemark{2}\\
\hline
Sublimation & $\circ$ & $\circ$ \\
Rotational breakup & $\circ$ & ? \\
Impact (cratering) & $\times$ & $\circ$ \\
Impact (shattering) & $\times$ & $\times$ \\
\hline\hline
P/2010 A2 & $\times$ & $\times$ \\
\hline
    \end{tabular}
  \end{center}
\tablenotetext{1}{Momentum conservation on the original body}
\tablenotetext{2}{Presence of the central body at the DEP (LF=DEP?)}
 \label{tab:mechanism}
\end{table}

\clearpage

\clearpage

\clearpage

\clearpage

\clearpage

\clearpage

\clearpage

\clearpage

\clearpage

\clearpage


\clearpage
\begin{thebibliography}{}

\bibitem[Agarwal et al.(2013)]{A13} Agarwal, J., Jewitt, D., \& Weaver, H.\ 2013, \apj, 769, 46 

\bibitem[Asada(1985)]{Asada1985} Asada, N.\ 1985, \jgr, 90, 12 

\bibitem[Asphaug et al.(1998)]{Asphaug1998} Asphaug, E., Ostro, S.~J., Hudson, R.~S., Scheeres, D.~J., \& Benz, W.\ 1998, \nat, 393, 437 

\bibitem[Birtwhistle et al.(2010)]{discovery} Birtwhistle, P., Ryan, W.~H., Sato, H., Beshore, E.~C., \& Kadota, K.\ 2010, Central Bureau Electronic Telegrams, 2114, 1 

\bibitem[Britt et al.(2002)]{Britt2002} Britt, D.~T., Yeomans, D., Housen, K., \& Consolmagno, G.\ 2002, Asteroids III, 485 

\bibitem[Burns et al.(1979)]{Burns1979} Burns, J.~A., Lamy, P.~L., \& Soter, S.\ 1979, \icarus, 40, 1 

\bibitem[Davis \& Ryan(1990)]{DR90} Davis, D.~R., \& Ryan, E.~V.\ 1990, \icarus, 83, 156 

\bibitem[Finson \& Probstein(1968)]{Finson1968} Finson, M., \& Probstein, R.\ 1968, \apj, 154, 327 

\bibitem[Fujiwara(1987)]{Fujiwara1987} Fujiwara, A.\ 1987, \icarus, 70, 536 

\bibitem[Giblin et al.(1998)]{Giblin1998} Giblin, I., Martelli, G., Farinella, P., et al.\ 1998, \icarus, 134, 77 

\bibitem[Hainaut et al.(2012)]{H12} Hainaut, O.~R., Kleyna, J., Sarid, G., et al.\ 2012, \aap, 537, A69 

\bibitem[Housen \& Holsapple(2015)]{HH15} Housen, K.~R., \& Holsapple, K.~A.\ 2015, Lunar and Planetary Science Conference, 46, 2894 

\bibitem[Holsapple \& Housen(2012)]{HH12} Holsapple, K.~A., \& Housen, K.~R.\ 2012, \icarus, 221, 875 

\bibitem[Housen \& Holsapple(1999)]{HH99} Housen, K.~R., \& Holsapple, K.~A.\ 1999, \icarus, 142, 21

\bibitem[Hsieh \& Jewitt(2006)]{HJ06} Hsieh, H.~H., \& Jewitt, D.\ 2006, Science, 312, 561 

\bibitem[Ishiguro et al.(2016)]{Ishiguro2016} Ishiguro, M., Kuroda, D., Hanayama, H., et al.\ 2016, \aj, 152, 169 

\bibitem[Ishiguro et al.(2013)]{Ishiguro2013} Ishiguro, M., Kim, Y.,  Kim, J., et al.\ 2013, \apj, 778, 19 

\bibitem[Ishiguro et al.(2011)]{I11} Ishiguro, M., Hanayama, H., Hasegawa, S., et al.\ 2011, \apjl, 741, L24 

\bibitem[Ishiguro(2008)]{Ishiguro2008} Ishiguro, M.\ 2008, \icarus, 193, 96 

\bibitem[Ishiguro et al.(2007)]{Ishiguro2007} Ishiguro, M., Sarugaku, Y., Ueno, M., et al.\ 2007, \icarus, 189, 169 

\bibitem[Jewitt et al.(2015)]{J15} Jewitt, D., Hsieh, H., \& Agarwal, J.\ 2015, Asteroids IV, 221 

\bibitem[Jewitt et al.(2013)]{J13} Jewitt, D., Ishiguro, M., \& Agarwal, J.\ 2013, \apjl, 764, L5 

\bibitem[Jewitt et al.(2010)]{J10} Jewitt, D., Weaver, H., Agarwal, J., Mutchler, M., \& Drahus, M.\ 2010, \nat, 467, 817 

\bibitem[Jutzi et al.(2010)]{Jutzi2010} Jutzi, M., Michel, P., Benz, W., \& Richardson, D.~C.\ 2010, \icarus, 207, 54 

\bibitem[Kim et al.(2012)]{K12} Kim, J., Ishiguro, M., Hanayama, H., et al.\ 2012, \apjl, 746, L11 

\bibitem[Kleyna et al.(2013)]{K13} Kleyna, J., Hainaut, O.~R., \& Meech, K.~J.\ 2013, \aap, 549, A13 

\bibitem[Marzari et al.(2011)]{Marzari2011} Marzari, F., Rossi, A., \& Scheeres, D.~J.\ 2011, \icarus, 214, 622 

\bibitem[Moreno et al.(2010)]{M10} Moreno, F., Licandro, J., Tozzi, G.-P., et al.\ 2010, \apjl, 718, L132 

\bibitem[Nakamura et al.(2015)]{Nakamura2015} Nakamura, A.~M., Yamane, F., Okamoto, T., \& Takasawa, S.\ 2015, \planss, 107, 45 

\bibitem[Nakamura \& Fujiwara(1991)]{N91} Nakamura, A., \& Fujiwara, A.\ 1991, \icarus, 92, 132 

\bibitem[O'Brien et al.(2011)]{5km} O'Brien, D.~P., Sykes, M.~V., \& Tricarico, P.\ 2011, Lunar and Planetary Science Conference, 42, 2665 

\bibitem[Okamoto \& Arakawa(2009)]{OA2009} Okamoto, C., \& Arakawa, M.\ 2009, Meteoritics and Planetary Science, 44, 1947 

\bibitem[Setoh et al.(2010)]{Setoh2010} Setoh, M., Nakamura, A.~M., Michel, P., et al.\ 2010, \icarus, 205, 702 

\bibitem[Snodgrass et al.(2010)]{S10} Snodgrass, C., Tubiana, C., Vincent, J.-B., et al.\ 2010, \nat, 467, 814 

\bibitem[Yanagisawa \& Itoi(1994)]{YI94} Yanagisawa, M., \& Itoi, T.\ 1994, 75 Years of Hirayama Asteroid Families:  The Role of Collisions in the Solar System History, 63, 243 

\end{thebibliography}
\end{document}